\documentclass[letterpaper,twocolumn,10pt]{article}
\usepackage{usenix-2020-09}

\usepackage{tikz}
\usepackage{amsmath}

\usepackage[backend=biber, style=numeric, maxnames=3, minnames=1]{biblatex}
\usepackage{float}

\usepackage[skip=0pt]{caption}

\usepackage{listings}
\definecolor{delim}{RGB}{20,105,176}
\definecolor{numb}{RGB}{106, 109, 32}
\definecolor{string}{rgb}{0.64,0.08,0.08}
\lstdefinelanguage{json}{
    numbers=left,
    numberstyle=\small,
    frame=single,
    rulecolor=\color{black},
    showspaces=false,
    showtabs=false,
    breaklines=true,
    postbreak=\raisebox{0ex}[0ex][0ex]{\ensuremath{\color{gray}\hookrightarrow\space}},
    breakatwhitespace=true,
    basicstyle=\ttfamily\small,
    upquote=true,
    morestring=[b]",
    stringstyle=\color{string},
    literate=
     *{0}{{{\color{numb}0}}}{1}
      {1}{{{\color{numb}1}}}{1}
      {2}{{{\color{numb}2}}}{1}
      {3}{{{\color{numb}3}}}{1}
      {4}{{{\color{numb}4}}}{1}
      {5}{{{\color{numb}5}}}{1}
      {6}{{{\color{numb}6}}}{1}
      {7}{{{\color{numb}7}}}{1}
      {8}{{{\color{numb}8}}}{1}
      {9}{{{\color{numb}9}}}{1}
      {\{}{{{\color{delim}{\{}}}}{1}
      {\}}{{{\color{delim}{\}}}}}{1}
      {[}{{{\color{delim}{[}}}}{1}
      {]}{{{\color{delim}{]}}}}{1},
}

\begin{document}

\date{}

\title{\Large \bf RapidPen: Fully Automated IP-to-Shell Penetration Testing with LLM-based Agents}

\author{
{\rm Sho Nakatani}\\
SecDevLab Inc.
} 

\maketitle

\begin{abstract}
  We present \textbf{RapidPen}, a fully automated penetration testing (pentesting) framework that addresses
  the challenge of achieving an initial foothold (\emph{IP-to-Shell}) without human intervention. Unlike prior
  approaches that focus primarily on post-exploitation or require a \emph{human-in-the-loop}, RapidPen
  leverages large language models (LLMs) to autonomously discover and exploit vulnerabilities, starting from
  a single IP address. By integrating advanced ReAct-style task planning (\emph{Re}) with retrieval-augmented
  knowledge bases of successful exploits, along with a command-generation and direct execution feedback loop
  (\emph{Act}), RapidPen systematically scans services, identifies viable attack vectors, and executes targeted
  exploits in a fully automated manner.
  
  In our evaluation against a vulnerable target from the Hack The Box platform, RapidPen achieved shell
  access within 200--400 seconds at a per-run cost of approximately \$0.3--\$0.6, demonstrating a
  \textbf{60\% success rate} when reusing prior “success-case” data. These results underscore the potential
  of truly autonomous pentesting for both security novices and seasoned professionals. Organizations
  without dedicated security teams can leverage RapidPen to quickly identify critical vulnerabilities,
  while expert pentesters can offload repetitive tasks and focus on complex challenges.
  Ultimately, our work aims to make penetration testing more accessible and cost-efficient,
  thereby enhancing the overall security posture of modern software ecosystems.
  \end{abstract}
  
  \section{Introduction}\label{sec:intro}
  
  Penetration testing (pentesting) typically begins with its most critical and challenging phase:
  \emph{initial infiltration} of a target system. Once an attacker—or in this case, a testing platform—gains
  an initial foothold, subsequent \emph{post-exploitation} tasks such as privilege escalation, credential
  theft, lateral movement, and data exfiltration become significantly more feasible. Although the initial foothold phase in penetration testing is challenging, preventing \emph{all} infiltration attempts is equally daunting, especially given the risks posed by zero-day exploits and social engineering. Consequently, it is essential to assess post-exploitation risks under
  the realistic assumption that a compromise may occur. The faster a testing process can confirm initial
  access, the more effectively it can allocate time to deeper post-exploitation stages before “running out
  of clock.”  
  
  Despite advances in automation, fully autonomous solutions for identifying initial compromise vectors remain elusive.
  In many cases, sophisticated pentesting still demands substantial human expertise, time, and cost. Recent
  advancements in large language models (LLMs) have driven progress in automating pentesting tasks, such as
  vulnerability scanning and post-exploitation. However, the \emph{initial-access} phase has received comparatively
  less attention. Existing approaches that incorporate LLMs often rely on a \emph{human-in-the-loop} to validate
  generated scans and exploits or to guide testing when ambiguities arise~\cite{pentestgpt2024}. While this
  approach may suit seasoned pentesters, it presents a significant barrier for software developers and system operators
  with limited security expertise, who may struggle to evaluate or refine the LLM’s recommendations. Moreover,
  prior research has identified two key challenges to full automation~\cite{wang2025unifiedmodelingframeworkautomated}:
  the vast search space of potential entry points and the highly target-specific nature of exploits.
  
  In this work, we focus on \emph{IP-to-Shell} testing: given only a target IP address, an autonomous system
  must obtain a shell without human intervention. Our goal is to develop a high-speed, low-cost
  solution that significantly simplifies penetration testing for both security professionals and non-specialists alike.
  
  \subsection*{Research Questions (RQs)}
  
  Comparing highly skilled human penetration testers with existing \emph{human-in-the-loop} systems~\cite{pentestgpt2024}, we hypothesize that two key design choices can facilitate autonomous, robust, and efficient initial infiltration:
  
  \begin{description}
    \item[RQ1:] \emph{Can reusing “success cases” (i.e., past experiences with successful scans and exploit paths) enhance the speed and reliability of initial-access automation?}
    \item[RQ2:] \emph{Does iterative command refinement—where the system analyzes failures and regenerates commands until successful—result in a higher probability of exploitation success?}
  \end{description}
  
  Building on these ideas, we propose an LLM-based pentesting agent, \emph{RapidPen}, which requires no human intervention beyond specifying a single target IP~\cite{rapidpen-demo}. We further broaden the scope of our investigation by posing the following research questions:
  
  \begin{description}
    \item[RQ3:] \emph{How does the time-to-compromise achieved by RapidPen compare to that of a skilled human pentester?}
    \item[RQ4:] \emph{How do the automation costs (in dollars per test) compare to manual penetration testing? Are they low enough to make automated solutions widely practical?}
  \end{description}
  
  \subsection*{Contributions and Scope}
  
  While our system is still in the early stages of development, we validate the feasibility of fully automated IP-to-Shell exploitation on a vulnerable target from the Hack The Box (\texttt{HTB}) platform~\cite{hackthebox}. Specifically, we achieve:
  \begin{itemize}
    \item A \textbf{60\% success rate} when leveraging past “success-case” data for the same class of vulnerability;
    \item Typical end-to-end compromise in \textbf{200--400 seconds};
    \item A per-run cost of only \textbf{\$0.3--\$0.6} to conduct a fully automated test.
  \end{itemize}
  
  These promising results support our hypotheses regarding the benefits of success-case knowledge (\textbf{RQ1}) and iterative command refinement (\textbf{RQ2}). We also provide a preliminary analysis relevant to \textbf{RQ3} and \textbf{RQ4}, comparing automated testing speed and cost with expert-driven testing.
  
  While RapidPen is designed to support organizations with limited security expertise, it is not exclusively intended for teams without dedicated security staff. We also envision its adoption by security teams and professional penetration testers, enabling them to offload straightforward assessments to RapidPen and focus their efforts on more complex, high-value testing scenarios. Like many other LLM-based penetration testing automation tools, RapidPen primarily achieves its objectives by flexibly leveraging existing knowledge. However, human expertise remains crucial for identifying novel vulnerabilities that are not yet documented or well understood.
  
  By providing a solution that benefits security-conscious organizations and industries with minimal prior expertise in penetration testing, we aim to enhance the overall security posture of modern software ecosystems.
  
  \textbf{Paper Outline.} 
  This paper is structured as follows: 
  Section~\ref{sec:background} reviews the fundamentals of penetration testing and the evolving landscape of AI-driven automation. 
  Section~\ref{sec:threatmodel} defines our threat model, scope, and assumptions, clarifying RapidPen’s operational boundaries. 
  Section~\ref{sec:design} presents the high-level architecture of RapidPen, including its ReAct-based modules and retrieval-augmented workflow. 
  Next, Section~\ref{sec:implementation} details our prototype implementation and core technical choices. 
  Section~\ref{sec:evaluation} describes our experimental setup and discusses the results of testing RapidPen on a vulnerable target. 
  Finally, Section~\ref{sec:conclusion} summarizes the key findings and outlines future directions for enhancing RapidPen’s capabilities and impact.
  
  \section{Background and Motivation}
  \label{sec:background}
  
  \subsection{Overview of Penetration Testing}
  Penetration testing (pentesting) is a structured process for identifying and validating vulnerabilities in
  systems and networks before malicious actors can exploit them. It typically involves multiple phases, which
  align with well-known frameworks such as the Penetration Testing Execution Standard (PTES)~\cite{ptes} or
  the MITRE ATT\&CK model~\cite{mitreattack}. Although different organizations may use slightly varying terminology, a common
  workflow includes:
  
  \begin{itemize}
    \item \textbf{Reconnaissance (Recon)} – Gathering preliminary information about the target, such as domain names,
          IP address ranges, and publicly available data. Effective reconnaissance can guide subsequent actions by identifying
          potential entry points.
    \item \textbf{Scanning (Enumeration)} – Conducting deeper, often automated probes on discovered services, ports,
          and configurations. Tools like \texttt{nmap} can identify vulnerabilities or anomalies.
    \item \textbf{Exploitation (Initial Access)} – Leveraging discovered weaknesses to gain unauthorized access. This
          phase is often the most challenging and high-stakes, as it determines whether an attacker can \emph{successfully}
          compromise the system.
    \item \textbf{Post-Exploitation} – Once an initial foothold is obtained, security testers (or adversaries) may
          escalate privileges, move laterally, and explore deeper layers of the environment to assess the impact
          of a breach.
  \end{itemize}
  
  Penetration testing plays a crucial role in cybersecurity: rigorous simulated attacks can expose complex
  weaknesses that static code analysis or automated scanners might overlook.
  By emulating real-world threats, pentesters help organizations prioritize remediation efforts and improve
  their overall security posture.
  overall security posture of modern software ecosystems.

  \subsection{LLM-Driven Automation in General}

  In recent years, large language models (LLMs) have rapidly advanced in both capability and scope, 
  enabling significant progress in automating a wide range of tasks, including natural language processing, 
  programming assistance, and more. Early breakthroughs include transformer-based architectures such as 
  \textbf{BERT} \cite{devlin2019bert}, \textbf{GPT-2/3} \cite{radford2019gpt2,brown2020gpt3}, 
  and \textbf{T5} \cite{raffel2019exploring}, which collectively demonstrated how pre-trained models 
  could perform text classification, summarization, and translation with minimal fine-tuning. 
  Subsequent models like \textbf{LaMDA} \cite{thoppilan2022lamda} and \textbf{GPT-4} \cite{openai2023gpt4} 
  have further increased parameter counts and the sophistication of emergent behaviors, 
  allowing for more complex and context-aware interactions.
  
  These advances have driven adoption across various application domains:
  \begin{itemize}
    \item \textbf{Text Summarization and Translation.} 
          LLMs trained on large corpora can generate concise summaries of lengthy documents 
          and translate text between multiple languages, often surpassing traditional systems 
          \cite{brown2020gpt3,raffel2019exploring}.
    \item \textbf{Code Generation and Debugging.} 
          Models such as Codex \cite{chen2021codex} can generate scaffolding code, unit tests, or entire functions 
          from natural language descriptions, accelerating software development and improving productivity. 
          Research also explores the use of LLMs for debugging and static analysis to identify potential software vulnerabilities.
    \item \textbf{Task Planning and Reasoning.} 
          Recent advancements integrate symbolic and factual reasoning with language models, facilitating 
          tasks such as chain-of-thought prompting \cite{kojima2022large} and multi-step planning \cite{react2022}. 
          These improvements enable structured decision-making in scenarios requiring multi-step execution and complex logic.
  \end{itemize}
  
  Since LLMs essentially learn a broad “prior” from large-scale text corpora, they can be adapted for novel 
  tasks through well-crafted prompts. This \emph{prompt engineering} paradigm significantly lowers 
  the barrier to automating domain-specific workflows, including cybersecurity-related tasks. Notably, 
  LLMs can parse tool outputs, synthesize commands, and adjust actions based on prior responses, 
  making them particularly well-suited for penetration testing scenarios requiring multi-step, context-aware orchestration.

  \subsection{Existing Research and Opportunities for Improvement}
  \label{sec:existingresearch}
  
  Recent research has explored the application of LLMs to automate various penetration testing tasks, from initial access to remediation.
  For example, \textbf{PentestGPT}~\cite{pentestgpt2024} introduces an LLM-based framework for guided exploitation
  using a task-tree architecture, while \textbf{PenHeal}~\cite{penheal2023} focuses on vulnerability discovery and
  mitigation strategies. Tools such as \textbf{BLADE}~\cite{blade2024} and \textbf{AutoAttacker}~\cite{autoattacker2024}
  extend automation into post-exploitation, and \textbf{Wintermute}~\cite{happe2023wintermute} highlights
  autonomous Linux privilege escalation.
  
  Despite these advancements, a key gap remains in achieving \emph{fast, fully automated initial infiltration}. To date,
  most approaches still rely on \emph{human-in-the-loop} validation or focus primarily on post-exploitation 
  rather than providing a high-speed, end-to-end framework for breaching a target. From a software development 
  and operations perspective, the critical questions are often, “Can my system be infiltrated, and how quickly
  can that happen?” Delivering an \emph{IP-to-Shell} workflow at practical speed and cost could provide significant 
  value to a broader audience, including security-conscious organizations and industries lacking dedicated security teams.
  
  In the following sections, we introduce an approach to address this need. By focusing on the initial-access phase
  and aiming for fully automated, low-cost, high-speed penetration testing, our work seeks to enhance overall 
  cybersecurity and enable a wider range of users to incorporate real-world adversarial testing into their
  development processes.

  \section{Threat Model and Problem Definition}
  \label{sec:threatmodel}
  
  \subsection{Threat Model}
  The RapidPen agent is assumed to have minimal prior knowledge of the target system:
  \begin{itemize}
    \item \textbf{Target IP Only.} The attacker (i.e., RapidPen) is provided only with the IP address of the
          machine under test, without additional configuration details or vulnerability disclosures.
    \item \textbf{Shell Acquisition.} In its current prototype, RapidPen exploits vulnerabilities using
          the Metasploit Framework~\cite{metasploit} (\texttt{msfconsole}) and considers a shell “obtained”
          once logs confirm that a reverse shell has been successfully established.
  \end{itemize}
  
  \subsection{Assumptions}
  RapidPen operates under the assumption that it can establish TCP connections to the target system’s IP address. 
  If necessary, an OpenVPN configuration file can be deployed within the RapidPen environment to enable
  VPN-based connectivity. Beyond these basic networking requirements, no additional external services or
  credentials are assumed.

\subsection{Scope and Limitations}
\begin{itemize}
  \item \textbf{Pre-Scanning Recon Excluded.}
        Passive reconnaissance steps, such as searching domain records or metadata leaks, are beyond the 
        scope of this study. Instead, we focus on active port scanning as the starting point.
  \item \textbf{No Post-Exploitation.} 
        RapidPen does not attempt privilege escalation or lateral movement once a shell is acquired.
  \item \textbf{No Web-Based Attacks.} 
        Although web vulnerabilities can serve as entry points, the current system does not address them. 
        Future work will explore extending RapidPen to support web exploits.
  \item \textbf{No UDP-Based Attacks.}
        This implementation is limited to TCP-based targeting. UDP-based exploits and scans are not considered in this study.
\end{itemize}

\section{Design Overview of RapidPen}
\label{sec:design}

In this section, we describe the overall architecture of our fully automated penetration testing framework,
referred to as RapidPen.
We adopt the \emph{ReAct}~\cite{react2022} paradigm, which consists of a \emph{Re} (task planning) module and an
\emph{Act} (command execution) module, both supported by specialized
\emph{retrieval-augmented generation (RAG)} \cite{rag2020} repositories.
Below, we detail the system architecture, how each module interacts, and how failures are handled.

\subsection{System Architecture}
\label{subsec:system-architecture}

\begin{figure}[H]
  \begin{center}
   \includegraphics[width=\linewidth]{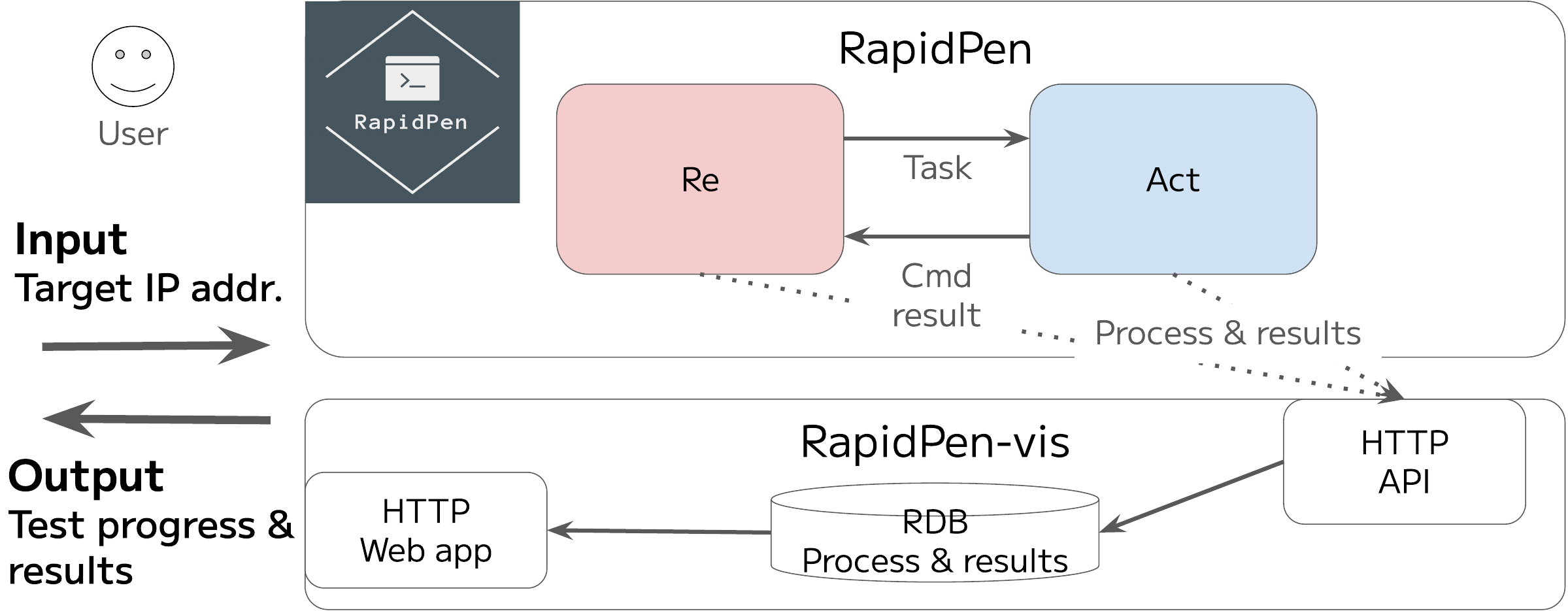}
  \end{center}
  \caption{\label{fig:rapidpen-architecture-overview} 
  The high-level architecture of RapidPen, illustrating user inputs and outputs, the \emph{Re} and \emph{Act} modules,
  and RapidPen-vis, a visualization tool for monitoring intermediate processes and final reports.
}
\end{figure}

Figure~\ref{fig:rapidpen-architecture-overview} provides a high-level overview of RapidPen’s core components.  

\begin{itemize}
  \item \textbf{Input}: The user provides the \emph{target IP address}.
  \item \textbf{Output}: RapidPen-vis displays penetration test progress (e.g., logs, discovered vulnerabilities) and generates the final reports, including the command used to obtain a shell.
  \item \textbf{Re and Act Modules}: These modules jointly implement the ReAct loop, coordinating tasks and executing commands.
  \item \textbf{RapidPen-vis}: A separate visualization tool for monitoring intermediate processes and final reports.~\cite{rapidpen-demo}
\end{itemize}

\subsection{PTT as a Core Data Model in the \emph{Re} Module}
\label{subsec:ptt}

In prior work on \textbf{PentestGPT}~\cite{pentestgpt2024}, the concept of a \textit{Pentesting Task Tree} (PTT) was introduced to structure the entire penetration testing process as an \emph{attributed tree}, where each node represents a task (e.g., port scanning, vulnerability testing, exploitation), and edges define the flow of reasoning or dependencies between tasks. The tree evolves dynamically as new tasks are generated, completed, or require backtracking due to partial failures.

\paragraph{PTT Definition (from PentestGPT).}
A PTT is essentially a labeled tree (or attributed polytree) with the following key elements:
\begin{enumerate}
    \item \textbf{Nodes (tasks)} with unique identifiers and optional child nodes.
    \item \textbf{Attributes} assigned to each node, such as task descriptions, current statuses, and relevant parameters.
    \item \textbf{Edges} representing parent-child relationships (e.g., subtask expansions) that structure the penetration testing workflow at multiple levels of detail.
\end{enumerate}

\paragraph{Our Extensions.}
We integrate the PTT as the \emph{core data model} in the \emph{Re} (reasoning) module to structure and coordinate tasks. In addition to the standard PentestGPT functionality, we introduce the following enhancements:
\begin{enumerate}
    \item \textbf{Environment Metadata.} Our PTT includes a dedicated \texttt{metadata} block capturing details about the penetration testing environment (e.g., attacker and target IP addresses, time stamps, test status).  
    \item \textbf{Act Results in Nodes.} Each task node maintains a history of command executions, including the executed command string, \texttt{exit\_code}, \texttt{exit\_class}, and a brief log summary. This allows for a clearer interplay between the \emph{Act} module outputs and the \emph{Re} module’s reasoning state.  
    \item \textbf{JSON-based I/O.} We consistently store and exchange the PTT in JSON format, ensuring that the LLM operates within a strict schema. This prevents ambiguity or “hallucination” when the LLM appends new tasks or updates existing nodes. A simplified JSON schema is provided in Listing~\ref{lst:ptt-sample}.
\end{enumerate}

\begin{lstlisting}[language=json,firstnumber=1,caption={Simplified PTT snippet with environment metadata, subtask structure, and Act results.},label={lst:ptt-sample}]
{
  "version": "2",
  "metadata": {
    "started_at": "2025-02-13T22:01:52Z",
    "finished_at": "2025-02-13T22:08:00Z",
    "status": "SUCCESS",
    "attacker": { "LHOST": "10.10.14.22" },
    "target": {
      "description": "HTB Blue machine",
      "RHOST": "10.10.10.40"
    }
  },
  "root": {
    "id": "1",
    "title": "Reconnaissance",
    "act_results": [{
        "command": "(omit)",
        "timeout_sec": 60,
        "exit_code": 0,
        "exit_class": "SUCCESS",
        "log_summary": "(omit)"
    }],
    "subtasks": [
      ...
    ]
  }
}
\end{lstlisting}

\subsection{Layered ReAct Modules in RapidPen}
\label{subsec:react-modules}

RapidPen’s execution logic follows the \emph{ReAct}~\cite{react2022} paradigm, where:
\begin{enumerate}
    \item \textbf{Re (Task Planning):} Monitors current logs, prior task outcomes, and “success-case” data to propose new tasks or exploit paths.
    \item \textbf{Act (Command Execution):} Issues commands to gather information or launch attacks. Upon receiving logs, the system refines or regenerates commands before feeding outcomes back to \emph{Re}.
\end{enumerate}

\begin{figure}[H]
  \begin{center}
    \includegraphics[width=\linewidth]{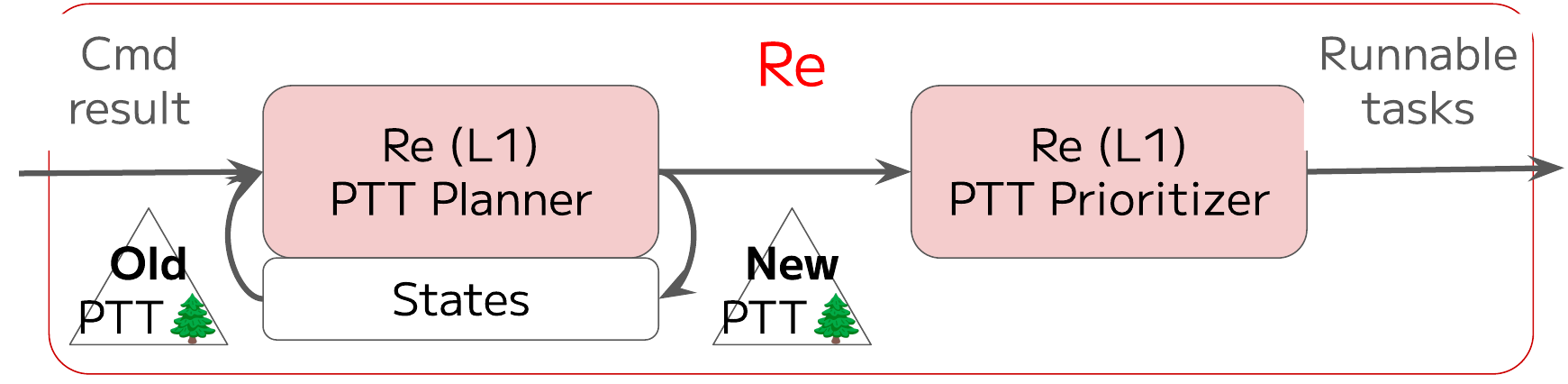}
  \end{center}
  \caption{\label{fig:re-architecture} 
  The \emph{Re} module in RapidPen consists of the \emph{Re (L1) PTT Planner} and \emph{Re (L1) PTT Prioritizer} submodules.
  The PTT Planner is responsible for expanding and maintaining the PTT tree, while the PTT Prioritizer determines
  the next task to execute.
  }
\end{figure}

\begin{figure}[H]
  \begin{center}
    \includegraphics[width=\linewidth]{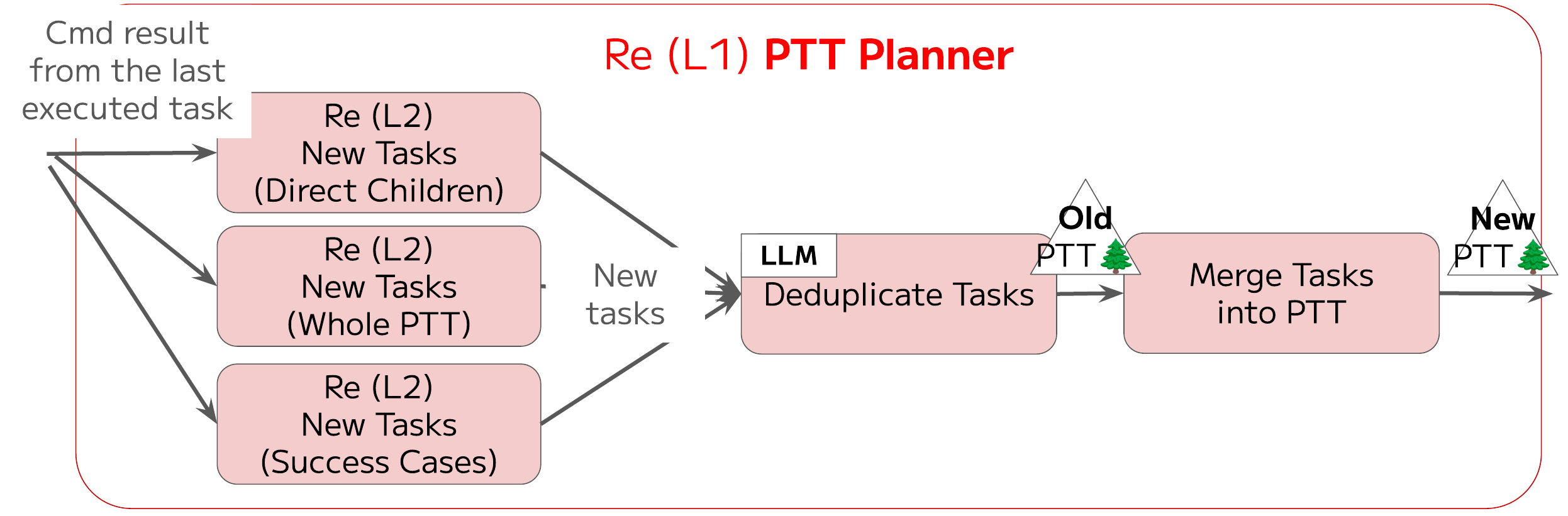}
  \end{center}
  \caption{\label{fig:re-l1-ptt-planner} 
  The \emph{Re (L1) PTT Planner} processes the command results from the last executed task to generate new tasks
  at Level 2 (L2) in the PTT. These tasks are deduplicated using an LLM-based approach before being merged
  into the PTT, updating it from an old to a new state.
  }
\end{figure}

\begin{figure}[H]
  \begin{center}
    \includegraphics[width=\linewidth]{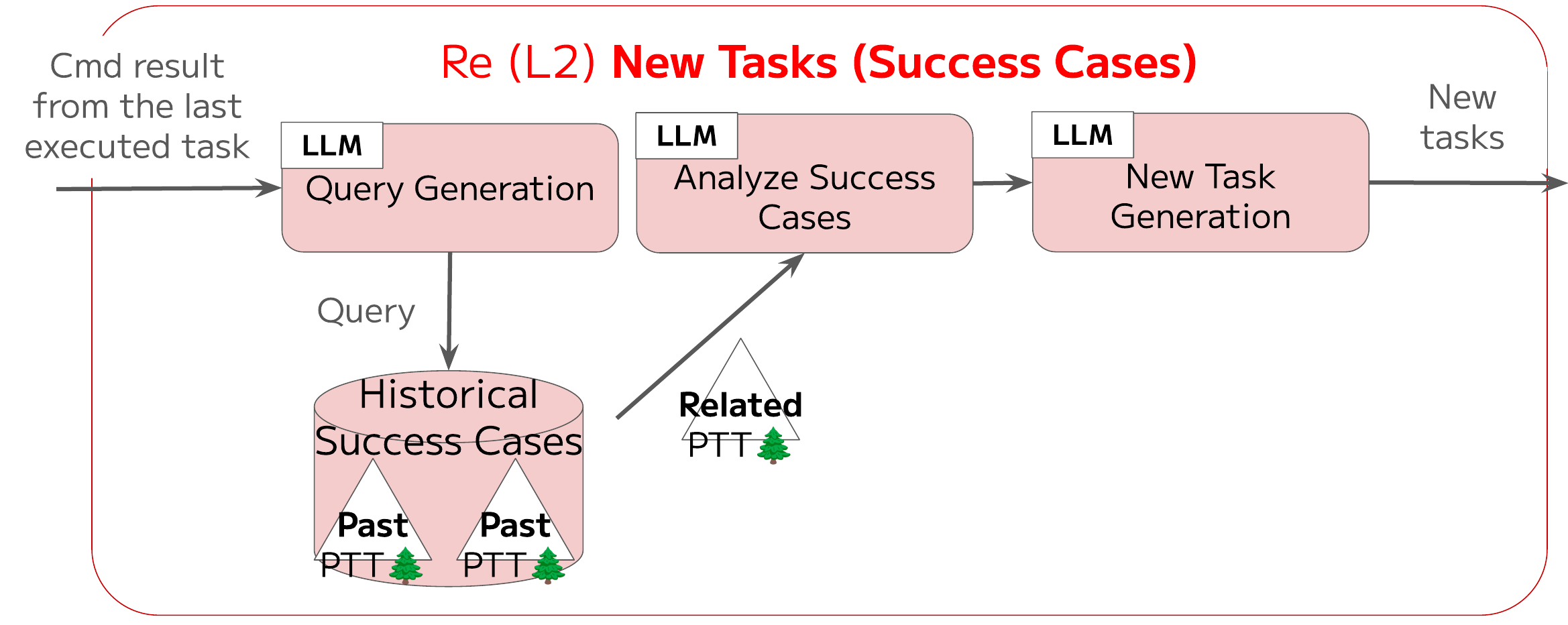}
  \end{center}
  \caption{\label{fig:re-l2-new-tasks-success-cases} 
  The \emph{Re (L2) New Task Generation} module generates new tasks based on historical success cases.
  The process begins by extracting command results from the last executed task.
  An LLM queries relevant historical success cases, which are then analyzed to extract key insights.
  Based on this analysis, new tasks are generated and integrated into the planning process.
  }
\end{figure}

\begin{figure}[H]
  \begin{center}
    \includegraphics[width=\linewidth]{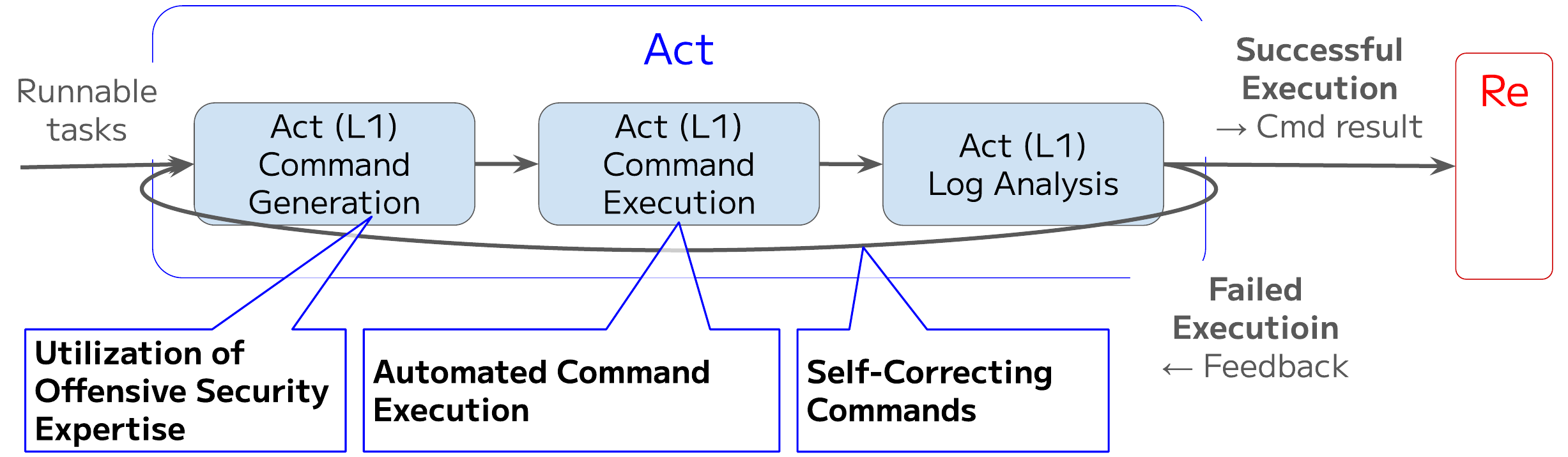}
  \end{center}
  \caption{\label{fig:act-architecture} 
  The \emph{Act (L1)} module processes runnable tasks through three key stages: command generation,
  execution, and log analysis. It leverages offensive security expertise to generate commands,
  automates execution, and applies self-correcting mechanisms when necessary. Successful executions
  produce a command result, while failed executions trigger a feedback loop for improvement.
  }
\end{figure}

Figures~\ref{fig:re-architecture}--\ref{fig:re-l2-new-tasks-success-cases} illustrate how the \emph{Re} module is divided into submodules, while Figure~\ref{fig:act-architecture} details the \emph{Act} module’s workflow.

\subsection{RAG for Offensive Security}
\label{subsec:rag-offensive}

While \emph{ReAct} provides a general “reasoning–acting” pattern, RapidPen enhances this approach with two specialized \emph{Retrieval-Augmented Generation (RAG)} repositories for domain-specific commands and proven exploit steps:

\begin{enumerate}
  \item \textbf{Act (L1) Command Generation RAG:}  
  A curated collection of 148 Markdown files from \textbf{HackTricks}~\cite{hacktricks}, primarily focused on “Network Services Pentesting” (e.g., SMB, FTP, SSH). These documents provide typical scan commands, exploit techniques, and enumeration strategies relevant to the initial-access phase. The Command Generation module references these documents to generate commands via the LLM.

  \item \textbf{Re (L2) New Tasks (Success Cases) RAG:}  
  PTTs in JSON format capturing successful pentesting sequences.  
  Currently, this dataset includes two PTTs for the Blue machine in Hack The Box~\cite{hackthebox}.  
  Each file outlines step-by-step instructions, from scanning to obtaining a shell.  
  The New Tasks (Success Cases) module generates a search query for the RAG based on the results of the most recent task execution.  
  It then analyzes the retrieved PTT output to generate effective subtasks.
\end{enumerate}

\subsection{Feedback Cycle in Act Module}
\label{subsec:act-feedback-cycle}

Figure~\ref{fig:act-architecture} illustrates the feedback loop within the \emph{Act} module, where command generation and execution are tightly coupled with log analysis and error handling. After executing each command, the system interprets the outcome (e.g., \texttt{SUCCESS}, \texttt{TIMEOUT}, \texttt{COMMAND\_NOT\_FOUND}) and determines whether to retry or escalate. The feedback loop follows these key policies:

\begin{enumerate}
  \item \textbf{Three-Strike Retry Limit.}  
  When commands are generated and executed in a cycle, the Log Analysis module evaluates the logs to determine if the result is conclusive.  
  If the command fails or does not produce sufficient evidence for further progress, the \emph{Act} module refines or regenerates the command and re-executes it.  
  This cycle repeats up to three times.  
  If no success is achieved after three attempts, RapidPen marks the corresponding task as \textit{failed} and reports this outcome to the \emph{Re} module.

  \item \textbf{Handling Timeouts.}  
  In some cases, command execution may hang indefinitely if the target server is unresponsive.  
  To prevent this, each command is assigned an initial timeout (e.g., 30~seconds).  
  When a \texttt{TIMEOUT} occurs, the next execution cycle begins with \emph{Act (L1) Command Generation} searching for a faster alternative command.  
  For example, it may replace an \texttt{nmap} port scan command with \texttt{rustscan} for quicker execution.  
  If no faster alternative exists, the system doubles the timeout threshold to allow more time for execution.

  \item \textbf{Handling Missing Commands or Files.}  
  Since \emph{Act (L1) Command Generation} references HackTricks and other sources, it may propose commands or reference files that are not available in the \emph{Act (L1) Command Executor} environment.  
  In such scenarios, \texttt{COMMAND\_NOT\_FOUND} or \texttt{FILE\_NOT\_FOUND} errors occur.  
  Upon detecting these, RapidPen employs a \textit{fail-fast} strategy: it terminates the current penetration test session and notifies the developer.  
  The rationale is that an external installation or environment fix is required before continuing, and automated retries would be ineffective.
\end{enumerate}

\section{Implementation}\label{sec:implementation}

This section describes the prototype implementation of RapidPen.  
While Chapter~\ref{sec:design} presented the overall design, here we focus on the specific tools, 
infrastructure, and configurations used to realize our fully automated pentesting workflow.

\subsection{Prototype Setup and LLM Usage}
Currently, RapidPen exists as a \textbf{prototype} implementation built on top of Dify\footnote{\url{https://dify.ai/}}.  
We run Dify locally to manage interactions with multiple Large Language Model (LLM) endpoints.  
Additionally, we integrate LangSmith\footnote{\url{https://smith.langchain.com/}} with Dify to precisely measure and monitor LLM invocation costs.  
This setup enables tracking of API calls, token usage, and associated costs under realistic testing conditions.

Our system exclusively employs OpenAI’s \texttt{gpt-4o}~\cite{gpt4o2024} as the underlying language model.  
Internally, we maintain \textbf{10 LLM instances} dedicated to the \emph{Re} module (task planning and reasoning) and \textbf{8 LLM instances} for the \emph{Act} module (command generation and log analysis).  
Initially, some prompts were adapted from \textbf{PentestGPT}~\cite{pentestgpt2024}; however, all prompts have since been replaced with original designs.

\subsection{RapidPen-vis}
For visualization and reporting, we provide \textit{RapidPen-vis}, consisting of:
\begin{itemize}
  \item \textbf{Server-Side:} A Python Flask application responsible for rendering real-time test logs and final pentest summaries.
  \item \textbf{Client-Side:} A lightweight \textbf{vanilla JavaScript} frontend that communicates with the Flask API to fetch and display pentesting progress graphically.
\end{itemize}
This interface allows operators to observe the automated exploit process, review execution logs, 
and track the overall state of penetration testing tasks.

\subsection{Custom Dify-Sandbox}
\label{subsec:custom-dify-sandbox}
Dify provides a secure Python execution environment called \textit{Dify-Sandbox}, which restricts system calls 
and external network access within a controlled Docker container. 
However, our \emph{Act (L1) Command Executor} requires broader system access to execute real-world pentesting commands. 
To address this limitation, we implemented a custom Docker image that maintains the same REST API interface 
as \textit{Dify-Sandbox}, but without restrictive sandbox policies. 
This customized container is integrated into our \texttt{docker compose} setup as a direct replacement 
for the official Dify-Sandbox. It processes the same API calls for command execution 
while permitting the necessary system calls and network interactions required for pentesting.

By leveraging this custom sandbox implementation, we maintain compatibility with Dify’s workflow 
and Python execution mechanism while removing constraints that would otherwise prevent valid pentesting operations. 
This dual approach ensures that our local environment remains modular and extensible, 
allowing for future experiments and improvements in pentest automation.

\section{Evaluation}\label{sec:evaluation}

This section presents preliminary experiments on the \textit{Legacy} machine from Hack The Box~\cite{hackthebox},
designed to validate RapidPen’s ability to establish an initial foothold (IP-to-Shell) 
in an early-stage prototype. Future work will extend these experiments to a broader set of targets.

\subsection{Objectives and Questions}
Our evaluation seeks to answer the following key questions:
\begin{enumerate}
    \item \textbf{Success Rate:} How often does RapidPen successfully achieve initial access (shell) on a known vulnerable machine?
    \item \textbf{Time and Bottlenecks:} How long does an average run take, and which modules in RapidPen consume the most time?
    \item \textbf{LLM Cost:} What is the cost of an automated penetration test in terms of LLM usage?
    \item \textbf{Behavioral Insights:} How does the feedback mechanism in the \emph{Act} module 
          (cf.~Section~\ref{subsec:act-feedback-cycle}) function in practice, and 
          what role does the \emph{Re (L2) New Tasks (Success Cases) RAG} (cf.~Section~\ref{subsec:rag-offensive}) 
          play in generating effective exploit paths?
\end{enumerate}

\subsection{Experimental Setup}
\paragraph{Target Machine (HTB Legacy).}
We selected the Hack The Box “Legacy” machine as our primary target. 
This machine features an older SMB server exposed on \texttt{tcp/445} with the \texttt{MS17-010} (EternalBlue) 
vulnerability, enabling remote code execution (RCE).

\paragraph{Attacker Environment.}
We executed the RapidPen orchestrator and \textit{RapidPen-vis} on a local MacBook~Pro (13-inch M2, 
24\,GB RAM, macOS Sequoia~15.3.1). The \textit{Dify}-based RapidPen orchestration runs in Docker containers, 
including our custom sandbox for actual command execution.

\subsubsection{Testing Two Configurations}
To evaluate the impact of \emph{Re (L2) New Tasks (Success Cases) RAG}, we conducted two sets of experiments:
\begin{itemize}
    \item \textbf{With Success Cases Enabled (Runs \#1--10):} The system had access to a stored PTT 
          reflecting successful exploitation steps on HTB “Blue” (which shares the MS17-010 vulnerability).
    \item \textbf{Without Success Cases (Runs \#11--20):} The system relied solely on scanning and 
          standard exploit references, without leveraging pre-recorded successful sequences.
\end{itemize}

For each run, we reset the environment, then launched RapidPen with a single target IP. 
We recorded the following metrics:
\begin{itemize}
    \item \textbf{Outcome:} \textit{Success} (obtained a shell) or \textit{Failure}.
    \item \textbf{\#Steps:} Number of PTT expansions initiated by the \emph{Re (L1) PTT Planner}, 
          from start to success/failure.
    \item \textbf{Elapsed Time:} Total wall-clock time from test initiation to termination.
\end{itemize}

\subsection{Results}
\subsubsection{Overall Success Rates and Timings}

\begin{figure*}[t]
  \begin{center}
    \includegraphics[width=\linewidth]{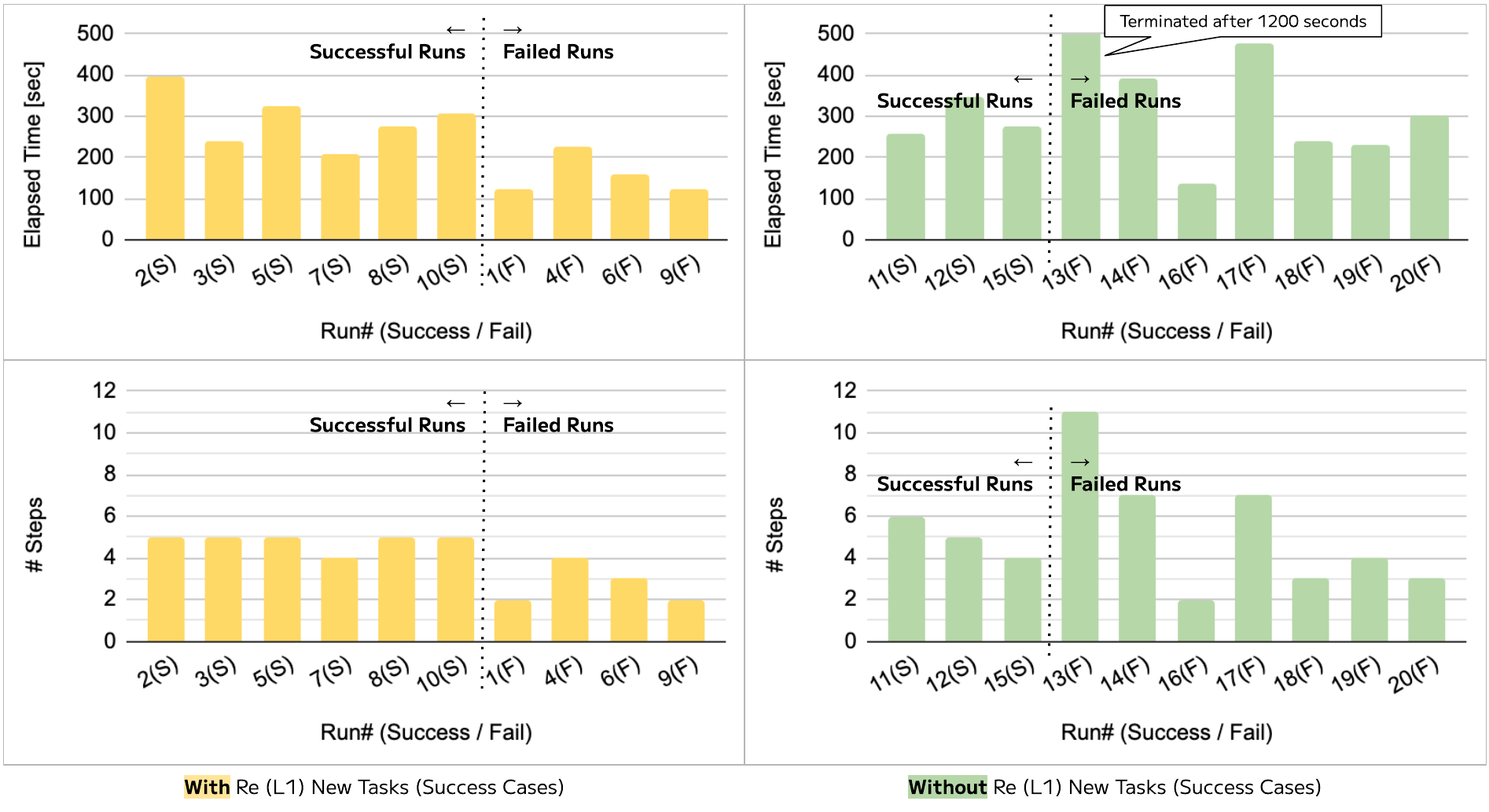}
  \end{center}
  \caption{\label{fig:experiment-basic} 
  Impact of \emph{Re (L2) New Tasks (Success Cases)} on Penetration Testing Efficiency.  
The left column (yellow) represents runs with \textbf{Success Cases enabled}, while the right column (green) represents runs \textbf{without Success Cases}.  
\textbf{Top Row:} Elapsed time (in seconds) per run. Runs with Success Cases tend to complete faster, while runs without them exhibit greater variance and some failures exceeding 1200 seconds (forcefully terminated).  
\textbf{Bottom Row:} Number of steps taken (PTT expansions) before success or failure. Runs without Success Cases often require more steps, indicating inefficient task selection.  
The vertical dashed line separates \textbf{successful} and \textbf{failed runs}, highlighting that the \textbf{failure rate is higher without Success Cases (3/10 success) compared to with Success Cases (6/10 success)}.
  }
\end{figure*}

\paragraph{With Success Cases (\#1--10).}
In the left column of Figure~\ref{fig:experiment-basic}, we observe that \textbf{6 out of 10} runs successfully achieved a shell on the Legacy machine.
Runs that failed tended to get stuck in repeated enumerations 
or timed out when \texttt{nmap} scanning did not produce conclusive results quickly. 
When the test succeeded, execution time ranged from 200--400\,seconds,
with a moderate correlation between the number of steps and elapsed time.

\paragraph{Without Success Cases (\#11--20).}
In contrast, the right column of Figure~\ref{fig:experiment-basic} presents a less favorable outcome.
The system succeeded in only \textbf{3 out of 10} runs. 
We also observed more outlier runs that either timed out after multiple scanning attempts 
or executed redundant exploit attempts. 
For instance, Run~\#13 followed an excessively long sequence of unsuccessful attack vectors.
Consequently, the average failure time was significantly higher, occasionally exceeding 400\,seconds. 
When an exploit succeeded, execution time was typically below 350\,seconds.

\subsubsection{Observations and Discussion}
The presence of \textit{Success Cases} significantly improved the success rate, 
as the Legacy machine shares the same SMBv1 vulnerability exploited by HTB ``Blue.'' 
Although the exact environment differs slightly, the fundamental MS17-010 exploit steps stored in the PTT 
closely align with the real target’s requirements. 
For more diverse vulnerabilities, we expect a lower direct transferability of RAG data; 
however, we anticipate that it will still accelerate the identification of effective enumeration and exploitation paths.

\begin{itemize}
    \item \textbf{Task Steps vs. Time:} Runs with fewer steps generally completed faster.
    \item \textbf{Failure Causes:}
          \begin{enumerate}
              \item Runs \#1, \#16, \#18, and \#20: \emph{Act (L1) Command Generation} produced \texttt{smbclient} commands with incorrect parameters, resulting in \texttt{COMMAND\_NOT\_FOUND}, \texttt{FILE\_NOT\_FOUND}, and \texttt{OTHERS} errors.
              \item Runs \#4, \#6, and \#17: \emph{Act (L1) Command Generation} generated inappropriate commands for the given tasks, leading to \texttt{OTHERS}, \texttt{FILE\_NOT\_FOUND}, and \texttt{COMMAND\_NOT\_FOUND} errors.
              \item Run \#9: \emph{Act (L1) Command Execution} failed to execute the \texttt{enum4linux} command on port 139. \emph{Act (L1) Log Analysis} classified it as an \texttt{OTHERS} error, triggering the fail-fast mechanism.
              \item Run \#13: The system continuously attempted to exploit port 139. After exceeding 1200 seconds, execution in Dify stalled.
              \item Run \#14: \emph{Re (L1) PTT Prioritizer} generated a hallucinated non-leaf task in the PTT, causing a validation error in the \emph{Act} module.
          \end{enumerate}
    \item \textbf{Future Generalization:} We plan to extend testing to machines with partially overlapping but not identical vulnerabilities to assess how well the success-case RAG generalizes.
\end{itemize}

\subsection{Module-Wise Time and Cost Breakdown}
To gain deeper insights, we instrumented the runs (particularly in the \textit{with Success Cases} scenario) 
to measure each module’s contribution to the total runtime and LLM costs. 

\begin{figure}[t]
  \begin{center}
    \includegraphics[width=\linewidth]{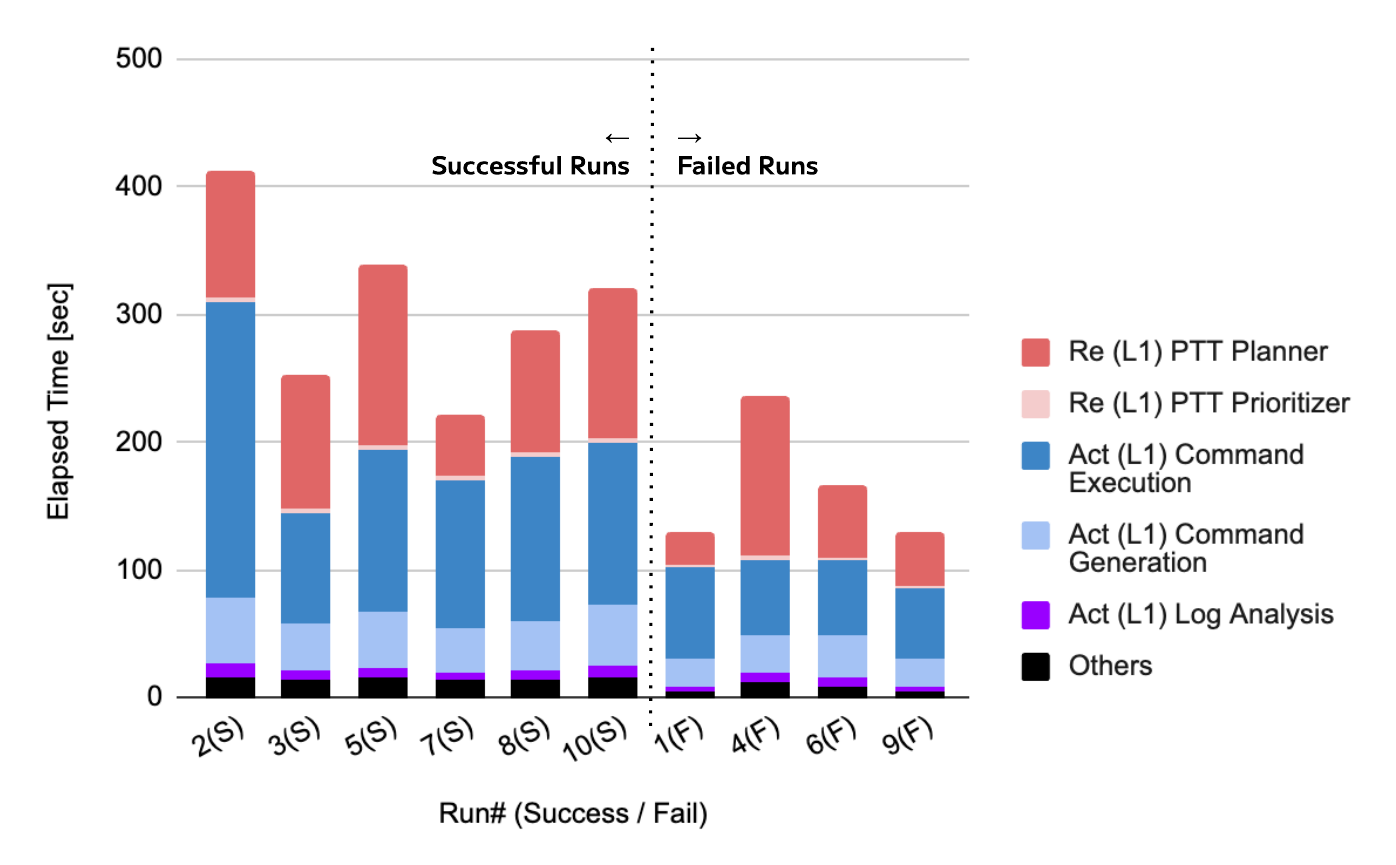}
  \end{center}
  \caption{\label{fig:experiment-time-breakdown} 
  Module-wise breakdown of elapsed time per run with \emph{Re (L2) New Tasks (Success Cases)} enabled.  
  Each bar represents the total execution time for a single run, with different colors indicating each module’s contribution.  
  \emph{Re (L1) PTT Planner} (red) accounts for a significant portion of the total time, followed by \emph{Act (L1) Command Execution} (blue).  
  }
\end{figure}

The runtime breakdown (see Figure~\ref{fig:experiment-time-breakdown}) indicates that \emph{Act (L1) Command Execution} 
contributes the most to total execution time, followed by \emph{Re (L1) PTT Planner}. 

\begin{figure}[t]
  \begin{center}
    \includegraphics[width=\linewidth]{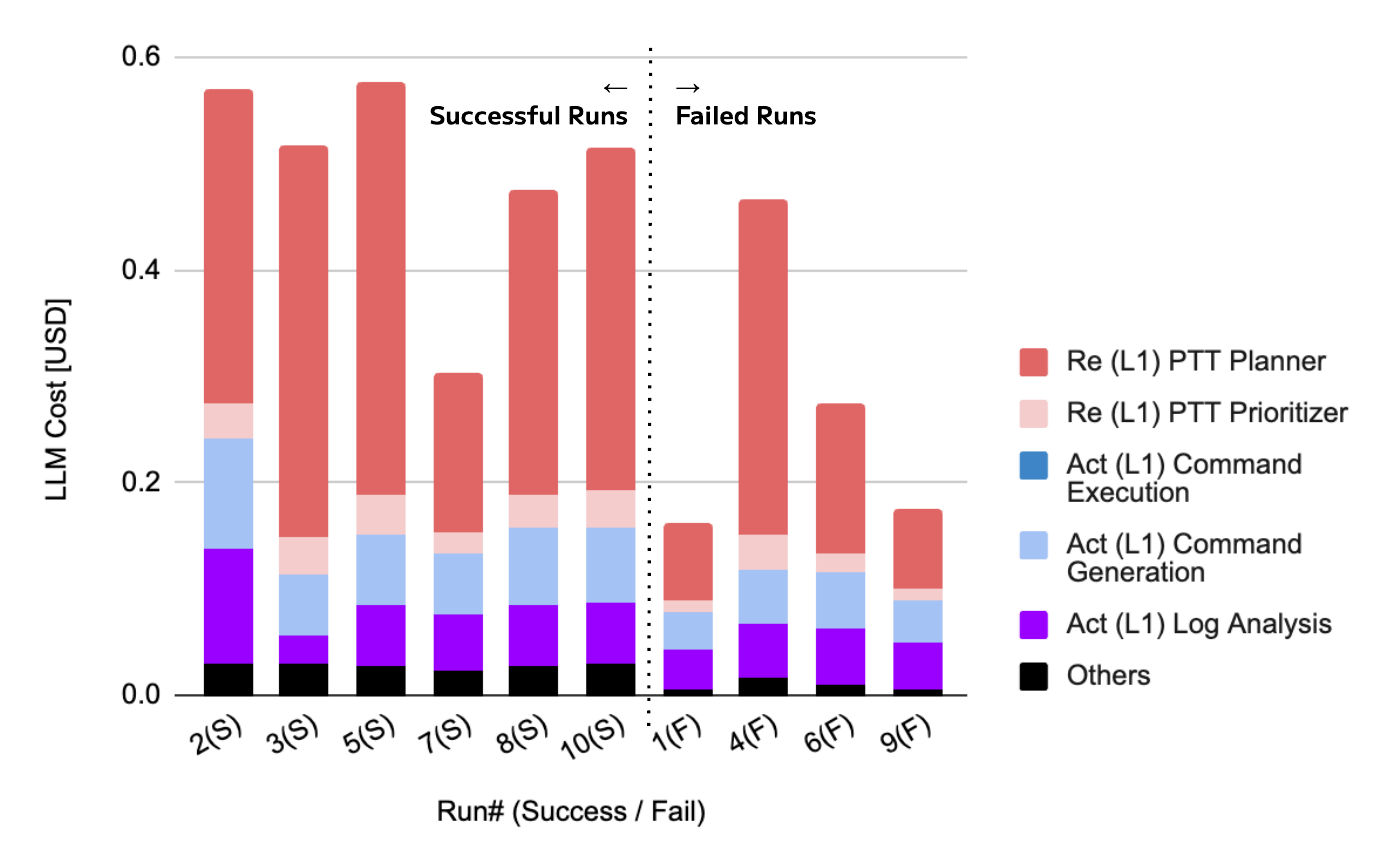}
  \end{center}
  \caption{\label{fig:experiment-cost-breakdown} 
  Module-wise breakdown of LLM cost per run with \emph{Re (L2) New Tasks (Success Cases)} enabled.  
  Each bar represents the total LLM cost (in USD) for a single run, with different colors indicating the cost contribution of each module.  
  \emph{Re (L1) PTT Planner} (red) incurs the highest cost, followed by \emph{Act (L1) Command Execution} (blue) and \emph{Act (L1) Log Analysis} (purple).  
  }
\end{figure}

Next, we analyze the LLM cost distribution (see Figure~\ref{fig:experiment-cost-breakdown}).  
\emph{Re (L1) PTT Planner} dominates the cost due to frequent PTT expansions and the overhead of merging newly generated 
subtasks from \emph{Re (L2) New Tasks (Success Cases)}.

The current cost and execution time are practical for targeted penetration testing scenarios.  
However, further optimization is possible by reducing large PTT inputs (which can sometimes exceed 14KB) to the LLM 
and improving error-handling mechanisms.

\subsection{Behavioral Insights}
\paragraph{Act Feedback Examples.}
As described in Section~\ref{subsec:act-feedback-cycle}, the \emph{Act} module attempts to recover from command failures by either adjusting parameters or switching to alternative tools. 
In multiple runs, we observed that when an \texttt{nmap} scan timed out, it was immediately replaced with \texttt{rustscan}. 
Additionally, when an exploit attempt using \texttt{msfconsole} timed out, the system generally did not find an alternative command and instead increased the timeout from 30 to 60 seconds before re-executing the command.

\paragraph{Role of Re (L2) New Tasks (Success Cases).}
Section~\ref{subsec:rag-offensive} introduced the \textit{New Tasks (Success Cases) RAG}, where RapidPen references a stored success PTT from HTB “Blue.”

\begin{lstlisting}[language=json,firstnumber=1,caption={Input to \emph{Re (L2) New Tasks (Success Cases)} in Legacy (last executed task)},label={lst:last-executed-task}]
{
  "id": "1.3.1.4",
  "title": "Enumerate services on port 445",
  "detail": "Use Nmap to enumerate the service running on port 445 on 10.10.10.4. Check for SMB vulnerabilities such as EternalBlue.",
  "status": "completed",
  "act_results": [
    {
      "command": "nmap -p 445 -vv -Pn --script= smb-vuln-cve2009-3103.nse, smb-vuln-ms06-025.nse, smb-vuln-ms07-029.nse, smb-vuln-ms08-067.nse, smb-vuln-ms10-054.nse, smb-vuln-ms10-061.nse, smb-vuln-ms17-010.nse 10.10.10.4",
      "timeout_sec": 30,
      "exit_code": 0,
      "exit_class": "SUCCESS",
      "log_summary": "TCP port 445 on 10.10.10.4 is open, and the following vulnerabilities have been detected: MS08-067 (CVE-2008-4250), MS17-010 (CVE-2017-0143)."
    }
  ],
  ...
}
\end{lstlisting}
  
When the last executed task from Listing~\ref{lst:last-executed-task} is passed to \emph{Re (L2) New Tasks (Success Cases)}, 
a query is generated for the \textit{New Tasks (Success Cases) RAG}:  
\texttt{"Metasploit SMB exploit port 445 empty credentials"}.  
This retrieves the stored success-case PTT from the HTB Blue machine (Listing~\ref{lst:blue-ptt}).

\begin{lstlisting}[language=json,firstnumber=1,caption={Success Case (PTT JSON) from the Blue machine},label={lst:blue-ptt}]
{
  "version": "2",
  "metadata": {
    "started_at": "2025-02-13T22:17:00Z",
    "finished_at": "2025-02-13T21:23:00Z",
    "status": "SUCCESS",
    "attacker": {
      "LHOST": "10.10.14.22"
    },
    "target": {
      "description": "HTB Blue machine",
      "RHOST": "10.10.10.40"
    }
  },
  "root": {
    ...
    {
      "id": "1.3.1.3",
      "title": "Enumerate services on port 445",
      "detail": "Use Nmap to enumerate the service running on port 445 on 10.10.10.40. Check for SMB vulnerabilities such as EternalBlue.",
      "status": "failed",
      "act_results": [
        {
          "command": "nmap -p 445 -vv -Pn --script= smb-vuln-ms17-010.nse 10.10.10.40",
          "timeout_sec": 30,
          "exit_code": 0,
          "exit_class": "SUCCESS",
          "log_summary": "Port 445 on 10.10.10.40 is open, and the Microsoft SMBv1 server has a remote code execution vulnerability (MS17-010, CVE-2017-0143)."
        },
        ...
      ],
      "subtasks": [
        {
          "id": "1.3.1.3.1",
          "title": "Exploit SMBv1 Vulnerability MS17-010",
          "detail": "Attempt to exploit the identified MS17-010 vulnerability on port 445 of 10.10.10.40 using Metasploit. Set RHOST to 10.10.10.40, RPORT to 445, and LHOST to 10.10.14.22.",
          "status": "completed",
          "act_results": [{
            "command": "msfconsole -q -x 'use exploit/ windows/ smb/ ms17_010_eternalblue; set RHOST 10.10.10.40; set RPORT 445; set LHOST 10.10.14.22; exploit'",
            "timeout_sec": 60,
            "exit_code": 0,
            "exit_class": "SUCCESS",
            "log_summary": "Port 445 on 10.10.10.40 is vulnerable to MS17-010. Windows 7 Professional 7601 Service Pack 1 x64 was identified, and a Meterpreter session was established."
          }],
          "subtasks": []
        }
      ]
  ...
}
\end{lstlisting}

Based on the success case from HTB Blue, the \emph{Re (L2) New Tasks (Success Cases)} LLM generated the following output:

\begin{lstlisting}[language=json,firstnumber=1,caption={Generated new tasks from Success Cases PTT.},label={lst:ptt-generated}]
{
  "Thought process (Step1)": "First, check whether a success-case task tree exists. The provided data contains a success-case task tree, so proceed to the next step.",
  "Thought process (Step2)": "Find a task in the success-case task tree similar to the most recent executed task. In this case, the latest task involves port 445, and the success-case task tree also contains detailed information about port 445. Specifically, task ID '1.3.1.3' and its subtasks are related to port 445. Based on these, generate new tasks.",
  "newTasks": [
    {
      "title": "Exploit SMBv1 Vulnerability MS17-010",
      "detail": "Attempt to exploit the identified MS17-010 vulnerability on port 445 of 10.10.10.4 using Metasploit. Set RHOST to 10.10.10.4, RPORT to 445, and LHOST to 10.10.14.22.",
      "origin": {
        "reason": "The previous check confirmed the presence of MS17-010 vulnerability on port 445, enabling an exploit attempt."
      }
    },
    {
      "title": "Verify System Access",
      "detail": "After exploiting the SMBv1 vulnerability, verify access to 10.10.10.4 by checking for a shell or other indicators of successful exploitation.",
      "origin": {
        "reason": "Verification is necessary to ensure that the exploit successfully provided access to the target system."
      }
    }
  ]
}
\end{lstlisting}

By leveraging the success case from the HTB Blue machine, which shares the same vulnerability, the system was able to generate appropriate tasks, demonstrating the effectiveness of using prior success cases for guided penetration testing.

\subsection{Summary of Findings}
Our preliminary evaluation indicates that RapidPen can achieve consistent IP-to-Shell exploits 
on a known vulnerable target:
\begin{itemize}
    \item \textbf{Success Rate:} 60\% with success-case RAG vs.\ 30\% without, across 10 trials each.
    \item \textbf{Time-to-Shell:} On average, 200--400\,seconds for successful runs.
    \item \textbf{LLM Cost:} Typically under \$0.60 per run, with the \emph{Re (L1) PTT Planner} module contributing the most.
\end{itemize}
Though limited to a single machine and vulnerability type, these results demonstrate 
the \emph{Re (L2) New Tasks (Success Cases)} approach’s potential and highlight 
the \emph{Act} module’s self-correcting behavior. 
We plan to broaden our scope with additional targets, diverse vulnerabilities, 
and larger user studies in future work.

\section{Discussion}
\label{sec:discussion}

\subsection{Benefits and Future Directions of the Act Feedback Mechanism}
The self-reliant feedback cycle implemented in the \emph{Act} module (Section~\ref{subsec:act-feedback-cycle}) significantly reduces the need for human intervention. As long as the tasks assigned by the \emph{Re} module are appropriate, the \emph{Act} module persistently re-generates and refines commands, interprets resulting logs, and explores alternative strategies when errors occur. This design choice allows RapidPen to continue progressing without manual oversight, enhancing its ability to achieve fully automated penetration testing.

However, the current fail-fast mechanism employed by the \emph{Act} module causes the entire process to terminate upon encountering specific errors, such as \texttt{COMMAND\_NOT\_FOUND}, \texttt{FILE\_NOT\_FOUND}, and \texttt{OTHERS}. While this approach prevents unnecessary retries and repeated failures, it can also abruptly halt the penetration test in cases where partial remediation—such as installing missing packages or updating outdated command syntax—would suffice. 

Future improvements should modify both \emph{Command Generation} and \emph{Command Execution} to address these errors dynamically. A more nuanced error-handling strategy should categorize failures, apply targeted retries or fixes, and reserve immediate termination for cases where it is strictly necessary. Such refinements would further enhance the system’s robustness and adaptability in real-world scenarios.

\subsection{Advantages and Limitations of Using Success Cases}
Our experiments indicate that RapidPen’s use of Success Cases accelerates exploit discovery when the target vulnerability closely matches those in previously recorded penetration tests. For example, referencing the MS17-010 exploit path from the “Blue” machine on Hack The Box (HTB) was effective against the “Legacy” machine, which shares a similarly vulnerable SMBv1 service. This demonstrates that reusing existing exploit sequences can streamline scanning and exploitation, leading to faster and more reliable outcomes.

However, handling scenarios where no relevant Success Cases exist remains an open problem. Zero-day vulnerabilities or configurations that have never been encountered may require more advanced reasoning beyond merely “copying” from past success. While our current approach leverages the LLM’s internal knowledge and a RAG-based repository, a more powerful framework for abstracting exploit techniques—enabling RapidPen to discover novel attack strategies—will be essential for addressing unknown threats. Designing and evaluating such a next-generation system is a critical step toward making automated pentesting broadly effective against new or rare vulnerabilities.

\subsection{Expanding the Attack Surface}
Although RapidPen currently achieves fully automated IP-to-Shell compromises, it does not yet address the post-exploitation phase. Privilege escalation, lateral movement, and deeper analysis of the compromised environment represent logical extensions for future work. In particular, tools like BLADE~\cite{blade2024} and AUTOATTACKER~\cite{autoattacker2024} already explore AI-assisted post-exploitation. Extending RapidPen to integrate with such frameworks could broaden its applicability, enabling more comprehensive, end-to-end assessments.

Another important direction involves web exploits, which are currently absent from the system. Web-based vulnerabilities often require specialized knowledge—ranging from injection techniques to authentication bypass methods—and may involve GUI-based testing beyond simple command-line interactions. Incorporating these capabilities would likely require RAG expansions to include relevant web exploitation knowledge bases and potentially adapt the \emph{Act} module to handle browser automation. Achieving the same degree of autonomy for web exploits poses additional research and engineering challenges.

\subsection{Ethical and Safety Considerations}
Although the user explicitly provides a target IP address to RapidPen, reducing the risk of scanning unrelated systems, the possibility of misuse cannot be ignored. Any automated exploit tool can be leveraged for malicious purposes if placed in the wrong hands or configured improperly. Future developments should focus on access control, rate-limiting, and formal usage policies—especially if the system transitions from a research prototype to a commercial or open-source deployment. Additionally, practical safeguards like monitoring logs, validating the legitimacy of the target environment, and enforcing strict network boundaries are pivotal for preventing inadvertent attacks against unauthorized hosts.

Overall, while RapidPen lowers the barrier for automated security testing, it underscores the need for responsible deployment practices. Addressing legal and ethical ramifications is essential to ensuring that the benefits of fully automated pentesting do not come at the expense of broader cybersecurity risks.

\section{Related Work}
\label{sec:relatedwork}

\subsection{LLM-Based Penetration Testing}

\subsubsection{PentestGPT – Task Tree-Driven AI Pentesting}
Recent research has explored using large language models (LLMs) to automate penetration testing. \textbf{PentestGPT} \cite{pentestgpt2024} is a notable example: it leverages an LLM (GPT-3.5/GPT-4) to guide the pentest process via a \textit{Pentesting Task Tree (PTT)} structure. Inspired by attack trees, the PTT decomposes engagements into sub-tasks (e.g., port scanning, service enumeration, exploitation), allowing the LLM to maintain context throughout testing. PentestGPT operates using three coordinated modules: a \emph{Reasoning} module (the "lead tester") that updates the task tree and determines next steps, a \emph{Generation} module (the "junior tester") that proposes specific commands, and a \emph{Parsing} module to summarize tool output.

While PentestGPT automates attack planning, it requires a human-in-the-loop to execute suggested commands and correct errors. Users must review and refine commands before execution, limiting its autonomy. Thus, PentestGPT functions more as a guided assistant rather than a fully autonomous pentesting tool.

\subsubsection{Other LLM-Driven Pentesting Tools}
Beyond PentestGPT, several emerging tools utilize LLMs for penetration testing, each focusing on different aspects of the workflow. These tools can be categorized as follows:

\paragraph{Tools that Automate Initial Access but Focus on Broad Vulnerability Scanning Rather than Speed}
\begin{itemize}
    \item \textbf{PenHeal} \cite{penheal2023} – an AI agent that operates without direct human involvement, designed to identify a broad range of vulnerabilities and propose mitigation strategies. Although the paper does not explicitly confirm automation of initial foothold attacks, it is possible that PenHeal’s capabilities overlap with RapidPen in terms of initial access. However, no evaluation is provided regarding the time and cost required to achieve initial access. In contrast, RapidPen focuses on demonstrating the most immediate security risk—namely, gaining unauthorized shell access as quickly as possible—before handing over control to established tools designed for post-exploitation. While RapidPen does not yet provide broad vulnerability coverage or automated remediation, incorporating such features remains an area for future exploration.
\end{itemize}

\paragraph{Tools That Focus on Post-Exploitation and Are Complementary to RapidPen}
\begin{itemize}
    \item \textbf{BLADE} \cite{blade2024} – \textbf{B}reaking \textbf{L}imits, \textbf{A}utomate \textbf{D}eep \textbf{E}xploitation – an AI-driven pentesting agent built on an autonomous agent framework (Microsoft’s AutoGen~\cite{autogen2023}). BLADE autonomously orchestrates exploitation tasks by leveraging external tools and dynamic script generation. For example, it uses pre-configured tools like LinPEAS for privilege escalation and John the Ripper for credential cracking to achieve deeper system compromise. Additionally, it includes agents for network scanning and lateral movement, showcasing how multi-agent AI systems can enhance penetration testing workflows.
    \item \textbf{AutoAttacker} \cite{autoattacker2024} – an LLM-guided system designed to implement automated “hands-on-keyboard” cyber-attacks in post-breach scenarios.
    \item \textbf{Wintermute} \cite{happe2023wintermute} – an LLM-driven Linux privilege escalation tool that evaluates model performance in fully automated exploit scenarios. It highlights strengths and weaknesses in autonomous security workflows, focusing on post-exploitation.
\end{itemize}

\subsection{Reinforcement Learning-Based Penetration Testing Approaches}

\textit{Deep reinforcement learning (RL)} has also been explored for autonomous pentesting. RL-based systems learn attack sequences by interacting with an environment and optimizing for successful exploits. Key contributions include:
\begin{itemize}
    \item \textbf{Hu et al.} \cite{hu2020deeprl} developed a deep RL framework for automated penetration testing, modeling scanning, exploitation, and lateral movement as a reinforcement learning problem.
    \item \textbf{Garrad and Unnikrishnan} \cite{garrad2023vanet} applied RL to vehicular ad-hoc network (VANET) penetration testing, demonstrating AI-driven attack sequence learning.
    \item \textbf{Liu et al.} \cite{liu2024hierarchical} proposed a hierarchical RL agent for large-scale network penetration, improving efficiency by splitting attack planning into multiple levels.
    \item \textbf{DeepExploit} \cite{takaesu2018deepexploit,deepexploit2019}, an early RL-powered pentesting tool integrated with Metasploit, demonstrated full automation of initial access but suffered from overfitting to training environments.
\end{itemize}
While RL-based pentesting can autonomously uncover attack paths, its major drawback is \textbf{poor generalization} beyond training data, requiring extensive retraining for new environments.

\subsection{Comparison with RapidPen}

\paragraph{Degree of Automation:} RapidPen is designed for full automation of initial access, requiring no human intervention once launched. This sets it apart from PentestGPT, which requires users to review and execute commands manually. In contrast, RL-based systems like DeepExploit require extensive training and tuning before deployment, making RapidPen a more practical choice for real-world pentesting with minimal setup overhead.

\paragraph{Scope of Initial Access Techniques:} RapidPen focuses on achieving unauthorized shell access as quickly as possible, covering a broad range of network and system-level exploitation techniques. Unlike PentestGPT, which primarily provides recommendations, RapidPen directly executes exploits. Meanwhile, tools like BLADE and AutoAttacker specialize in post-exploitation rather than initial access, making them complementary rather than competing solutions.

\paragraph{Usability for Non-Experts:} RapidPen is explicitly designed for usability by non-experts, enabling security assessments without deep penetration testing expertise. Unlike PentestGPT, which still requires expert validation of generated commands, RapidPen autonomously performs the entire attack process. Additionally, tools like Autonomous Web Exploitation target a different domain (web applications), leaving gaps in usability for broader infrastructure pentesting.

Overall, RapidPen distinguishes itself by combining \textbf{full automation, speed, and accessibility}. It provides a \textbf{highly practical and deployable solution} for automated initial access testing, making it a valuable tool for security practitioners and organizations looking to assess their exposure to real-world attack scenarios.

\section{Conclusion and Future Work}
\label{sec:conclusion}

In this paper, we introduced \textbf{RapidPen}, a fully automated penetration testing framework aimed at achieving an \emph{IP-to-Shell} compromise without human intervention. By combining ReAct-style task planning with retrieval-augmented exploit knowledge and iterative command generation/execution loops, RapidPen systematically scans for vulnerabilities and exploits them, demonstrating promising results on a vulnerable Hack The Box target. Our evaluation shows that RapidPen can achieve shell access within minutes at a modest cost, even in its current prototype form.

\subsection{Summary of Contributions}
\begin{itemize}
  \item \textbf{Proposal and Implementation.} We described the design of RapidPen’s modular
        \emph{Re} and \emph{Act} subsystems, highlighting how each leverages large language
        models and curated knowledge repositories.
  \item \textbf{Empirical Evaluation.} Preliminary experiments on a known vulnerable target
        demonstrated up to a \textbf{60\% success rate} for shell acquisition within
        \textbf{200--400 seconds}, with a per-run cost of approximately \textbf{\$0.3--\$0.6}.
  \item \textbf{Key Insights.} Our results highlight how reusing “success cases” and employing
        self-correcting command loops significantly enhance the reliability and efficiency of automated
        pentesting.
\end{itemize}

\subsection{Future Directions}

\paragraph{Expanding the Scope.}
Although RapidPen is currently designed for TCP-based initial access, we plan to extend its capabilities
in several areas:
\begin{itemize}
  \item \textbf{Web and UDP Attacks.} Expanding support for web-based exploits, including injection and
        authentication bypass techniques, and exploring UDP-based vulnerabilities as logical
        next steps.
  \item \textbf{Beyond Initial Access.} Integrating passive reconnaissance and post-exploitation workflows 
        (e.g., lateral movement, privilege escalation) by interfacing RapidPen with complementary automated or manual tools.
\end{itemize}

\paragraph{Refining the Current Implementation.}
Our short-term development focuses on improving reliability and performance within RapidPen’s existing
scope:
\begin{itemize}
  \item \textbf{Robust Error Handling.} Strengthening the \emph{Act (L1) Command Execution} and \emph{Act (L1) Log Analysis} pipeline
        to prevent premature termination from unexpected failures and clarify when retries or alternative commands
        are appropriate.
  \item \textbf{Optimized PTT Input to LLM.} Pruning irrelevant fields or tasks when feeding
        \textit{Pentesting Task Tree} (PTT) JSON data to the LLM to reduce context size, thereby
        increasing speed and lowering costs, particularly in \emph{Re (L1) PTT Planner}.
\end{itemize}

\paragraph{Execution Modes and Trade-offs.}
To improve success rates, we plan to introduce an “auto-retry” mode, where RapidPen automatically re-runs
failed test attempts. This feature will be user-configurable, allowing for a balance between execution time,
cost, and a higher probability of success.

\paragraph{Towards Real-World Deployment.}
We aim to make RapidPen accessible to a broader audience—whether through commercial offerings or as an
open-source project—so that software teams and security professionals alike can benefit from automated
initial-access testing. At the same time, we must design appropriate safeguards to minimize the risk of
misuse and ensure that RapidPen is deployed exclusively in legitimate, authorized environments.

\subsection{Closing Remarks}
By focusing on \emph{IP-to-Shell} automation, our work provides both security novices and experts with a
powerful tool for quickly identifying critical exposures. We envision that RapidPen’s foundation in
LLM-driven planning and execution can serve as a stepping stone toward a new class of intelligent, extensible
offensive security tools. As RapidPen matures, we hope it will stimulate further research into
collaborative workflows between humans and AI agents, ultimately strengthening the security posture
of modern software ecosystems.

\printbibliography

\end{document}